\documentclass[preprint,showpacs,aps]{revtex4}
\usepackage{epsfig}
\usepackage{graphicx}
\usepackage{dcolumn}
\usepackage{bm}
\begin{document}
\preprint{APS/123-QED}
\title{Full bandstructure calculation of second harmonic generation susceptibility of $\alpha-LiIO_{3}$ crystal}
\author{Yung-mau Nie}
\email{ymnie@phys.sinica.edu.tw}
\affiliation{Institute of Physics, Academia Sinica, 128 Sec. 2, Academia Rd, Nankang, Taipei 115, Taiwan, R.O.C.}
\date{\today}
\begin{abstract}
The present work performs full bandstructure calculations to
investigate the structural effect and the transition mechanisms of
the second harmonic generation (SHG) susceptibility of the
$\alpha-LiIO_{3}$ crystal. The anomalous inconsistency of associated
experimental data of the SHG susceptibility tensor is elucidated to
be dominated by the structural effect especially on the topology of
$O$-atoms. On the manipulation of the structural effect, the
modification of SHG susceptibility using an external pressure is
simulated. The calculations of SHG susceptibility tensor are
completed at finite frequencies and the static limit. The comparison
with the experiments is also incorporated. On the analysis of the transition mechanisms, the
inter-band transition is determined to entirely dominate the whole
SHG susceptibility at the static limit; however, the effect of the
intra-band motion is revealed to be as important as that of the
inter-band transition at finite frequencies.
\end{abstract}
\maketitle
\section{\label{sec1}Introduction}
\indent Recently the research of second harmonic generation (SHG) of
semiconductors has revived, mainly due to the boost on technologies
of the associated second-order phenomena: the electro-optic effect
and the optical rectification. The former, widely utilized in the
electro-optic modulators to tune the intensity of optical signals,
plays a crucial role in switching and routing optical broadband
networks of the up-to-date telecommunications.\cite{Kippelen} The
latter, usually viewed as an inverse operation to the former, is an
important mechanism for generating pulsed tera-hertz radiations,
whose applications include the cellular level imaging, the
semiconductor and high-temperature characterization, the chemical
and biological sensing, etc. \cite{Ferguson}
In fact, the SHG susceptibility of semiconductors dominates the efficiencies of both aforementioned effects. Thus, the investigation of its physical insights not only guides the modification of frequency doubling in the laser engineering, but also inspires the innovation of technologies of the associated second order phenomena.\\
\indent The alpha-phase lithium iodate ($\alpha-LiIO_{3}$) crystal
is a famous SHG material for the merits: frequency doubling of the
widely used radiation of Yd:YAG, a high nonlinear coefficient, a
weak temperature dependence of refractive index, a high damage
threshold, etc. \cite{Buesener} Therefore this crystal was
commercialized via the long-term studying in experiments.
\cite{Nath,Nash,Campillo,Jerphagnon,Otagurs,Levine155,Choy,Eckardt,Roberts,Borsa,Alford,Singh-up,Okada,Pearson,Liu,Kato,Buesener} Moreover, very recently it has been generalized to the nano-structural systems for the application of nonlinear optical waveguide. \cite{Teyssier1,Teyssier2} However, the published experimental data of the SHG susceptibility tensor of this crystal exhibits an anomalous inconsistency, such as the variation form $-12$ to $10$ pV/m about the $d_{31}$ component at static limit. \cite{Singh} This is in contrast to the consistency in measurements of the linear dielectric constant.\cite{Singh} So far this anomaly is still an open question. To addressing on this feature, the current investigation performs full bandstructure calculations basing on the implement of First-principles simulations. In fact, according to the best knowledge, the present work is the first quantum-level simulation of the SHG susceptibility for this crystal.\\
\indent
The present study reveals the structural effect as one of the dominating factors to the aforementioned inconsistency of experimental data for this system. Since the macroscopic polarization constituted by the dipoles of unit cells defines the SHG susceptibility, the study of this effect is motivated by the variation of dipoles for the considerable disagreement between the reported structural refinements \cite{Svensson,Rosenzweig} and the existing metamorphosis of this polar crystal \cite{Rosenzweig,Emiralieb}. In fact, the present results highlight this structure effect especially on the influence of the topology of $O$-atoms to the $\alpha-LiIO_{3}$ crystal. This argument is analogous to the recently reported ideology that modifies the nonlinear optic susceptibility of polar-ordered systems via the field-manipulating structural twist.\cite{Gubler} In order to attempt further to take advantage of the structural effect, the current study simulates the modification of the SHG susceptibility of the $\alpha-LiIO_{3}$ crystal via loading an external pressure. On the other hand, the organic polymer-based systems, the mainstream of nonlinear optic materials in the next generation, \cite{Kippelen} are not only usually in the form of the assembly of curled fibers, but also with the high-flexibility property fitting in the pressure manipulation. Thus, the results of structural effect emphasized in present article is very heuristic to the engineering about the SHG susceptibility of those organic materials.\\
\indent The resolution on the mechanisms of SHG susceptibility of
the $\alpha-LiIO_{3}$ crystal is also dedicated in this work.
According to the previous analysis of bandstructure of the
$\alpha-LiIO_{3}$ crystal by the '$\textbf{ k} \cdot p$' expansion
\cite{Nie}, the importance of the portion pertaining to the
intra-band transition has predicted, so the adequate theory should
distinctively incorporate the contribution of this mechanism. Hence,
the length-gauge formalism developed by Sipe et al.
\cite{Sipe1,Sipe2,Sipe3,Sipe4,Sipe5} is adopted for the full
bandstructure calculation, which is established within the framework
of the time-dependent perturbation for the interaction of the
independent particle with the field of long-wavelength. Starting
from the remedy of the zero-frequency divergence in the previous
works \cite{Aspnes, Levine}, Sipe and Ghahramani included the sum
rule of the periodic parts of the Bloch functions \cite{Lax} to
vanish the coefficient of the divergent term with respect to the
crystals with filled bands.\cite{Sipe1} Due to the incorporation of
the sum-rule, the effect of intra-band motion was explicitly
included and extracted to be defined as independent terms from the
portion of the inter-band transition. At the same quantum level to
calculate the SHG susceptibility of crystals, other methods include:
the application of "2n+1" theorem of the action functional of
time-dependent DFT \cite{Corso}, the bond-charge model
\cite{Levine1970}, and the application of the Berry phase in
macroscopic polarization \cite{Vanderbilt}.
They are able to yield accurate results to consist with the experimental measurements; however, the physical insights about the transition mechanism provided from those ways are very limited.\\
\indent
The components of the inter-band transition, the intra-band transition, and the modulation of the inter-band part by the intra-band transition are respectively calculated in wide frequency spectra. Owing to most of all published data at the static limit, the present study especially addresses on the simulations at this point, in which the adopted formalism \cite{Rashkeev,Sipe5} is specified to analytically satisfy the Kleinman symmetry \cite{Kleinman, Franken}. At finite frequencies, the Kleinman symmetry breakdowns that is not suited for SHG of cw laser radiation.\cite{Buesener} Thus, the associated experimental works were seldom performed before. However, the linear electro-optic effect and the optical rectification are always operated in the finite frequency regime, especially at high frequencies. Then the present results at finite frequencies are good reference to the researches of the associated second order phenomena.\\
\indent
Since the strongly dynamic charge-redistribution stimulated in the high-power lasers had been emphasized \cite{Marques}, thus once the validity of the perturbation scheme was suspected due to the considerable disagreement with the experimental measurements for semiconductors, i.e. $GaAs$ \cite{Sipe3}. Later the argument was postulated to resolve the doubt that the strong screening effect of the semiconductors deteriorates the fault of ignoring the non-local effect in the local-density approximation (LDA) of the density functional theory (DFT) to influence the performance of full bandstructure calculation.\cite{Lambrecht,Levin43,Aulbur} Thus, the modification of the scissor operator was proposed in order to improve the accuracy of the calculation. \cite{Sipe3,Rashkeev,Nastos}
However, such a non-local problem in semiconductors is determined to disappear in the present system, according to the resulting band-gap values consisting with the experimental data. In fact, the present results manifests the perturbation method still to be credible for simulating the SHG susceptibility of the $\alpha-LiIO_{3}$ crystal; however, except the above non-local issue, to consider the consistency with the experiments needs to further incorporate other subtle factors, i.e. the structural effect proposed in the present investigation. Thus, the conclusions of this article are very worthy to refer as applying the same scheme to other wide-gap oxides-series nonlinear inorganic crystals.
\section{\label{sec2} COMPUTATION AND FORMALISM}
\subsection{FIRST-PRINCIPLES CALCULATIONS}
\indent
The calculations are performed within the scheme of the density functional theory (DFT) with the generalized gradient approximation (GGA), in which the exchange-correlation potential functional is parameterized by Perdew et al. \cite{PBE}.
 Both ways of all-electron calculation, the projector augmented wave (PAW) method \cite{Blochl1,Kresse1} and the modified full-potential linearzed augmented plane wave (FLAPW) method \cite{Blochl} are respectively applied by the implements of VASP \cite{Kresse2,Kresse3} and WIEN2k packages \cite{Blaha1, Blaha2, Blaha3, Madsen}. To simulate the equilibrium of Hellmann-Feynman forces, the PAW method is applied to determine the energetically optimal structure parameters, in which the atomic relaxation is achieved by the conjugate-gradient scheme. The FLAPW method is used for the full bandstructure calculation, due to its derivative quantity, electric field gradient (EFG), to be useful on the analysis of structure refinements. Therein the core and the valence states are respectively calculated relativistically and semi-relativistically. The valence states include the $2s$-states for the $Li$-atom; the $2s$- and the $2p$-states for the $O$-atom, and the $4d$-, the $5s$-, and the $5p$-states for the $I$-atom. The muffin-tin radii are set to be $1.8$, $1.95$, $1.5$ {\AA} for the $Li$-, the $I$- and the $O$-atom, respectively. Inside each muffin-tin sphere, the expansion of associated Legendre polynomials for spherical harmonics of the wave function is truncated at $l=10$. The value of parameter $RK_{max}$ was set to be $8.0$. To improve the linearization in the semi-core regime, extra local orbit are incorporated for the low-lying states: the $4d$-states of the $I$-atoms and the $2s$-statess of the $Li$- and the $O$-atoms.
\subsection{FORMALISM}
\indent
The presently adopted formalism at finite frequencies is composed of three components: the inter-band transitions $\chi^{abc}_{2inter}(-2\omega;\omega,\omega)$, the intra-band transitions $\chi^{abc}_{2intra}(-2\omega;\omega,\omega)$, and the modulation of inter-band terms by intra-band terms $\chi^{abc}_{2mod}(-2\omega;\omega,\omega)$, as follows \cite{Sipe4}
\begin{widetext}
\begin{eqnarray}
\chi_{2}^{abc}(-2\omega;\omega,\omega)&=& \chi_{2inter}^{abc}(-2\omega;\omega,\omega)+ \chi_{2intra}^{abc}(-2\omega;\omega,\omega)+ \chi_{2mod}^{abc}(-2\omega;\omega,\omega), \label{eqn:SHG-FF}\\
\chi_{2inter}^{abc}(-2\omega;\omega,\omega)&=&\frac{e^{3}}{\hbar^{2}}\int\frac{d{\bf k}}{4\pi^{3}}\{\Sigma_{nml}\frac{r_{nm}^{a}\{r_{ml}^{b}r_{ln}^{c}\}}{(\omega_{ln}-\omega_{ml})}[\frac{2f_{nm}}{(\omega_{mn}-2\omega)}+\frac{f_{ml}}{(\omega_{ml}-\omega)}+\frac{f_{ln}}{(\omega_{ln}-\omega)}]\}; \label{eqn:SHG-inter}\\
\chi_{2intra}^{abc}(-2\omega;\omega,\omega)&=&\frac{e^{3}}{\hbar^{2}}\int\frac{d{\bf k}}{4\pi^{3}}\left[\Sigma_{nml}\omega_{mn}r_{nm}^{a}\{r_{ml}^{b}r_{ln}^{c}\}\right. [\frac{f_{nl}}{\omega_{ln}^{2}(\omega_{ln}-\omega)}-\frac{f_{lm}}{\omega_{ml}^{2}(\omega_{ml}-\omega)}]\nonumber\\
&-&8i\Sigma_{nm}\frac{f_{nm}r_{nm}^{a}\{\Delta_{mn}^{b}r_{mn}^{c}\}}{\omega_{mn}^{2}(\omega_{mn}-2\omega)}
+\left. 2\Sigma_{nml}\frac{f_{nm}r_{nm}^{a}\{r_{ml}^{b}r_{ln}^{c}\}(\omega_{ml}-\omega_{ln})}{\omega_{mn}^{2}(\omega_{mn}-2\omega)}\right];\label{eqn:SHG-intra}\\
\chi_{2mod}^{abc}(-2\omega;\omega,\omega)&=&\frac{e^{3}}{2 \hbar^{2}}\int\frac{d{\bf k}}{4\pi^{3}}\left[\Sigma_{nml}\frac{f_{nm}}{\omega_{mn}^{2}(\omega_{mn}-\omega)}\right.(\omega_{nl}r_{lm}^{a}\{r_{mn}^{b}r_{nl}^{c}\}-\omega_{lm}r_{nl}^{a}\{r_{lm}^{b}r_{mn}^{c}\})\nonumber \\
&-&\left.i\Sigma_{nm}\frac{f_{nm}\Delta_{nm}^{a}\{r_{mn}^{b}r_{nm}^{c}\}}{\omega_{mn}^{2}(\omega_{mn}-\omega)}\right]. \label{eqn:SHG-mod}
\end{eqnarray}
\end{widetext}
It is worthy to note that the formula of equation (\ref{eqn:SHG-mod}) has corrected the original wrong typing of the equation (B3) printed in the reference \cite{Sipe4}.
The result of the imaginary part is yielded by the analytic solution of the pole $\omega+i\delta$ for all $\delta \rightarrow +0$.
Subsequently, the result of real part is determined by the solved imaginary portion via the Kramers-Kronig relation.
Here $f_{nm}$ is defined as $f_{n}-f_{m}$, which $f_{n}$ deotes the Fermi occupation function to be $1$ ($0$) for the occupied (empty) state.
The bracket $\{r_{mn}^{b}r_{nl}^{c}\}$ is defined as $\{r_{mn}^{b}r_{nl}^{c}\} \equiv \frac{1}{2}(r_{mn}^{b}r_{nl}^{c}+ r_{mn}^{c}r_{nl}^{b})$.
Since the momentum matrix element $p_{nm}^{a}$ is much easier obtained from the resulting periodic Bloch function than the position matrix element $r_{nm}^{a}({\bf k})$, then the latter is given via the following relationship \cite{Sipe2, Sipe3}
\begin{equation}
r_{nm}^{a}({\bf k})=\frac{p_{nm}^{a}({\bf k})}{im\omega_{nm}({\bf k})},\label{eqn:x-p}
\end{equation}
where a finite $r_{nm}^{a}({\bf k})$ only exists for $\omega_{n}({\bf k}) \neq \omega_{m}({\bf k})$.
The matrix element $\Delta_{nm}^{a}$ has the definition, $\Delta_{nm}^{a}\equiv v_{nn}^{a}-v_{mm}^{a}$ , where $v_{nn}^{a}$ is the intra-band velocity matrix element.
The problem with the unphysical divergence previously mentioned by Rashkeev at al \cite{Rashkeev}, also occurs with the present study in computing the $r_{mn}^{a}$ when the bands $n$ and $m$ are nearly degenerate.
Referring to the previous treatment \cite{Rashkeev}, here also set $r_{nm}^{a}=0$ when $\hbar \omega_{nm} \le \epsilon$ with the small cutoff $\epsilon$ of $5 \times 10^{-4} Ryd$ to remove the influence of this problem.\\
\indent
At the static limit, the formulism \cite{Rashkeev,Sipe5} analytically fulfilling the Kleinman symmetry is adopted, respectively expressing the inter-band and the intra-band transition as follows
\begin{widetext}
\begin{eqnarray}
\chi_{2}^{abc}(0;0,0)&=&\chi_{2inter}^{abc}(0;0,0)+ \chi_{2intra}^{abc}(0;0,0);\label{eqn:SHG-st}\\
\chi_{2inter}^{abc}(0;0,0)&=& \frac{e^{3}}{\hbar^{2}}\int\frac{d{\bf k}}{8\pi^{3}}\Sigma_{nml}\frac{ r_{nm}^{a}\{r_{ml}^{b}r_{ln}^{c}\}}{\omega_{ml}\omega_{ln}\omega_{mn}}(\omega_{m}f_{nl}+\omega_{n}f_{lm}+\omega_{l}f_{mn});\label{eqn:SHG-st-inter} \\
\chi_{2intra}^{abc}(0;0,0)&=& \frac{ie^{3}}{4\hbar^{2}}\int\frac{d{\bf k}}{8\pi^{3}}\Sigma_{nm}\frac{f_{nm}}{\omega_{mn}^{2}} \left[(r_{nm}^{a}r_{mn}^{b})_{;c}^{-}\right.+ \left.(r_{nm}^{a}r_{mn}^{c})_{;b}^{-}+(r_{nm}^{b}r_{mn}^{c})_{;a}^{-}\right], \label{eqn:SHG-st-intra}
\end{eqnarray}
\end{widetext}
where $(r_{nm}^{a}r_{mn}^{b})_{;c}^{-}\equiv r_{nm}^{a}r_{mn;c}^{b}- r_{nm;c}^{a}r_{mn}^{b}$, and $r_{mn;c}^{b}$ is defined as \cite{Sipe5}
\begin{equation}
r_{mn;c}^{b}=-\frac{(r_{mn}^{b}\Delta_{mn}^{c}+r_{mn}^{c}\Delta_{mn}^{b})}{\omega_{mn}}-\frac{1}{\omega_{mn}}\Sigma_{l}(v_{ml}^{b}r_{ln}^{c}-r_{ml}^{c}v_{ln}^{b}).\nonumber
\end{equation}
\indent
The special-point sampling method of Monkhorst and Pack \cite{Monkhorst} is applied for the integration of all above formula in Brillouin-zone. To consider the remarkable fluctuation within the {\bf k}-space to the inter-band energy $\omega_{mn}$ and the momentum matrix $p_{nm}$, to achieve the convergence of the integration is only by means of a fine {\bf k}-mesh.
Hence, the samplings of $12 \times 12 \times 11$ and the $23 \times 23 \times 21$ meshes were tested for the convergence in advance.
Both give a difference of about $10$ percent, comparable to the error range in the experiments, for the calculations of the equation (\ref{eqn:SHG-st}) and even better consistency of the calculations at finite frequencies. Thus, the former is used in the present calculations to give creditable estimations.\\
\section{\label{sec3} ANALYSIS FOR STRUCTURE REFINEMENTS}
\begin{figure}[htbp]
\resizebox{0.9\textwidth}{!}{%
  \includegraphics{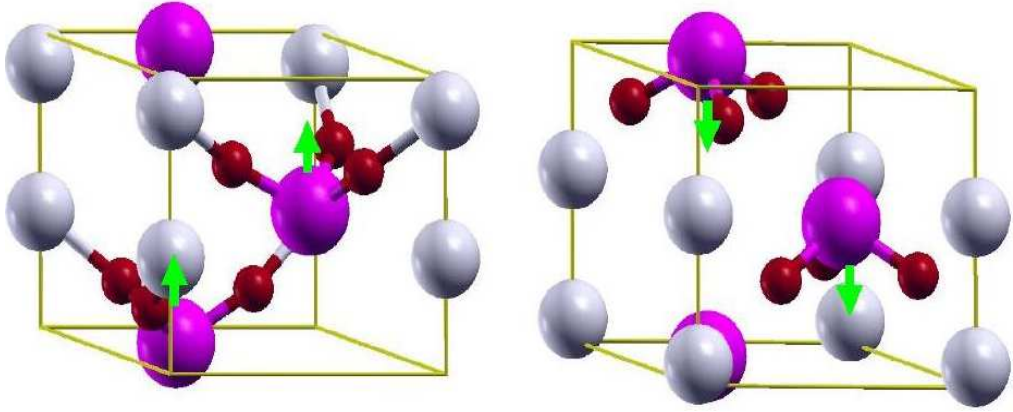}}
\caption{\label{fig:structure}(Color online) The different structure
refinements of the $\alpha-LiIO_{3}$ crystal. The structure (A) and
(B) are given by the x-ray diffraction results \cite{Rosenzweig} and
\cite{Svensson}, sketched as left and right figures, respectively.
The refinements of the energetically optimal structure and the
distorted structure in simulating the effect of an external pressure
along the $z$-axis, are similar to that of structure (B). The
dipoles of the $IO_{3}^{-1}$ anions are sketched by the green arrow
symbols, where the ones of structure (B) are defined as positive.
The $I$-atom, the $O$-atom and the $Li$-atom are sketched as the
purple, the red, and the grey spheres.}
\end{figure}

\begin{figure}[htbp]
\resizebox{0.75\textwidth}{!}{%
  \includegraphics{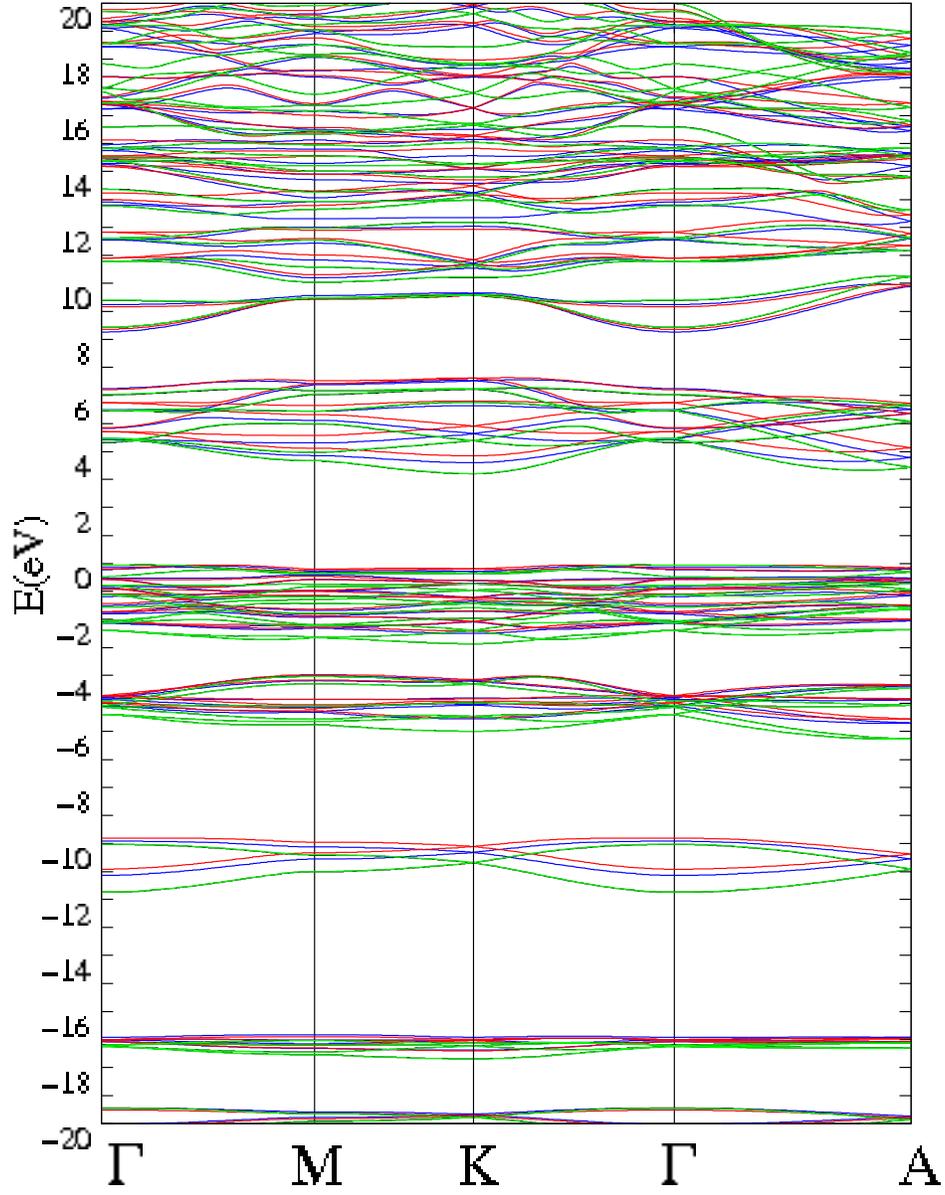}}
\caption{\label{fig:bandstructure}(Color online) The bandstructure for the simulated structures. The blue, the red, the black, and the green lines respectively depict the results of the structure (A), (B), (C), and (D).}
\end{figure}

\indent
The unit-cell parameters of two experimental results and the energetically optimal structure, determined by the equilibrium of Hellmann-Feynman forces, are respectively tabulated in the TABLE~\ref{tab:structure}. In order to simulate the pressure effect to the SHG susceptibility of the $\alpha-LiIO_{3}$ crystal, the distorted structure with a $z$-dimension reduced by $1$ millionth of the original length with respect to the above optimal structure is calculated; this equivalently simulates to load an average pressure of $\sim 3770$ Gpa normal to $xy$-plane to the original system according to the calculation. After the shrinking, its status at the static equilibrium is simulated by the atomic relaxation, whose parameters are also exhibited in the TABLE~\ref{tab:structure}. The structure refinements of simulated cases are illustrated in the FIG.~\ref{fig:structure}. The associated results on the bandstructure, the value of band gap $E_{g}$, and the principle component of EFG tensor $V_{zz}$, are illustrated in the FIG.~\ref{fig:bandstructure} and the TABLE ~\ref{tab:structure}, respectively.\\
\indent
The resulting bandstructures of all calculated structures only differ in the inter-band energies, subjected to the relative atomic positions in the unit cell, and agree with each other in the curvature, due to the similarity in the lattice constants. Especially, the energetically optimal structure and the associated distortion almost have a same bandstructure. Thus, according to the tabulated atomic positions with respect to the fixed $I$-atom, the different topology of $O$-atoms is deduced to dominate the change of bandstructure near the edges by comparing with the results of structure (A) and (B), since the previous study \cite{Nie} has revealed the highly ionic feature of the $Li$-atom to cause little participation in the states near the Fermi-level. The maximum and the minimum of the resulting band-gap values are $3.934$ and $3.297$ eV. The former consists with the resonant onset indicated in the absorption \cite{Regel, Xu, Galez} and the transmission spectrum \cite{Nash}; however, the latter is comparatively closed to the recently experimental result ($\approx 3$eV) \cite{Gaffar}. Because of the resulting values of $E_{g}$ consisting with the experimental results, the effect of non-local defect of DFT-LDA can be ignored in the present calculations.\\
\indent The resulting $V_{zz}$ is given by the term $l=2$, and $m=0$
in the full-potential expansion \cite{Blaha1,Blaha2} of the FLAPW
calculation. In experiments, it is determined by the resonant
frequency of nuclear spin quadruple moment for the nuclei with a
spin quantum number $I \geq 1$. \cite{Christiansen} The associated
experimental value of the $V_{zz}$ for the $I$($Li$)-atom is $-52.8$
($0.075$), determined by the cited measurements: the resonant
frequency $151.27 \times 10^{6}$ ($36.4\times 10^{3}$) Hz, and the
asymmetry parameter $0.017$ ($0$) in the transition from the nuclear
magnetic moment $1/2$ to $3/2$ \cite{Baisa} ($1/2$ to $3/2$
\cite{Minamisono}); the quadruple moment $-0.789$ barn \cite{Pyykko}
($0.04$ barn \cite{Voelk}).
In general, the results of the $I$- and the $Li$-atom agree with the experimental values; it implies the accuracy of the present FLAPW calculations, since the EFG itself can be viewed as a criterion on capturing the detail electronic structure of the system.\\
\indent
According to the similarities of the resulting $E_{g}$ and $V_{zz}$ for all simulated structures and the agreement with the experiments, these structures should be equally possible to be operated in the lab. Thus, the inconsistency of results defined by them can be considered as the nature in the experiments.\\
\section{\label{sec5} SECOND HARMONIC GENERATION RESULTS}
\subsection{Static Limit}
\indent The resulting independent nonzero components: $d_{31}$ and
$d_{33}$; the associated experimental results are shown in
TABLE~\ref{tab:d31} and  ~\ref{tab:d33}. The component $d_{31}$ can
be directly measured in the crystal itself by the phase-matching
method
\cite{Nath,Eckardt,Roberts,Borsa,Alford,Nash,Campillo,Jerphagnon,Otagurs,Levine155,Choy};
however, the $d_{33}$ has to be measured by the 'wedge' method
\cite{Nash, Jerphagnon,Choy,Roberts} via another crystal, whose SHG susceptibility needs to be known. In order to clinically reduce the erroneous
judgement from any external factor, the comparison with the
experiments more emphasis on the component of $d_{31}$ in the
present study. Herein the tabulated portion of the visual-hole
(visual-electron) process in the inter-band transition is defined by
the Eq.(\ref{eqn:SHG-st-inter}), in which the transitions completes
via two valence (conduction) states. On the other hand, due to the
lack of consistent experimental data to refer for the
$\alpha-LiIO_{3}$ crystal, the test on a well-known system for the adopted full bandstructure calculation of the SHG susceptibility becomes necessary, which can be viewed as a kind of 'alignment'. Thus, the
well-studied $GaAs$ system was selected. Basing on the same FLAPW
method to compute the bandstructure and the scissors operation
suggested by Nastos et
al\cite{Nastos} to deal with the non-local defect of DFT-LDA, the resulting value $185$ pm/V is rather closed to the experimental result $168$ pm/V and the previously simulated results collected in the reference \cite{Rashkeev}. In general, the resulting magnitude of the SHG susceptibility of the $\alpha-LiIO_{3}$ crystal is at the same order of the experimental data.\\
\indent
The importance of the structural effect to the SHG susceptibility of the $\alpha-LiIO_{3}$ crystal reflects on the significant variation of the results for different structures on both $d_{31}$ and $d_{33}$ components. First of all, basing on the previous study on the bandstructure by the '$\textbf{ k} \cdot p$' expansion \cite{Nie}, the agreement on the curvatures of obtained bandstructures for different structures illuminates the consistency of the resulting momentum matrix elements $p_{mn}^{a}$, so the influence of $p_{mn}^{a}$ to the structural effect can be excluded. Thus, according to the analysis of bandstructures for all structures in Sec.~\ref{sec3}, this structural effect are deduced to be caused by the change on the inter-band energies $\omega_{mn}$, which is dominated by the topology of $O$-atoms, previously also mentioned in Sec.~\ref{sec3}. Furthermore, the negative result of the structure (A) is in contrast to the positive results of the other structures, can be attributed to its unique inverse $O$-topology among all calculated structures, depicted in the FIG.~\ref{fig:structure}. This inverse $O$-topology feature causes a negative dipole of the $IO_{3}^{-}$ anion to compose a negative macroscopic polarization of the system. Such a polarization is completely equivalent to that caused by the presently obtained negative susceptibility in the same system. Thus, the present investigation illuminates the structural effect, especially on the topology of the $O$-atoms, dominating the magnitude and the sign of the SHG susceptibility tabulated in the TABLE~\ref{tab:d31} and \ref{tab:d33}. This conclusion can explain why the inconsistency of the data appears to the different specimens in spite of repeating measurements via the same operation conditions in the lab.\cite{Otagurs,Choy,Alford}\\
\indent
On the respect of manipulating the structural effect via an external pressure, the magnitude of the modification is about $10$ percent to the $d_{31}$ according to the present simulation. Although this scheme seems neither efficient on the low enhancement nor economic on the requested pressure being as high as $3770$ Gpa; however, it is still an effective way to modified the SHG susceptibility of the $\alpha-LiIO_{3}$ crystal.\\
\indent
On the resolution of the transition mechanisms, the contribution of the inter-band transition distinctively prevails that of the intra-band transition to both components. Moreover, the analysis on the compositions of the inter-band transition reveals the weights of the visual-hole and the visual-electron process differing between $d_{31}$ and $d_{33}$ components, even though the results of both components given by the same bandstructure. This feature highlights the extreme anisotropic character of the SHG susceptibility of the $\alpha-LiIO_{3}$ crystal. In addition, in fact the structural effect significantly changes the proportions of the inter-band and the intra-band transitions, and those of the visual-hole and the visual-electron components in the inter-band transition.
\subsection{Finite Frequency Regime}
\begin{figure}[htbp]
\resizebox{0.45\textwidth}{!}{%
  \includegraphics{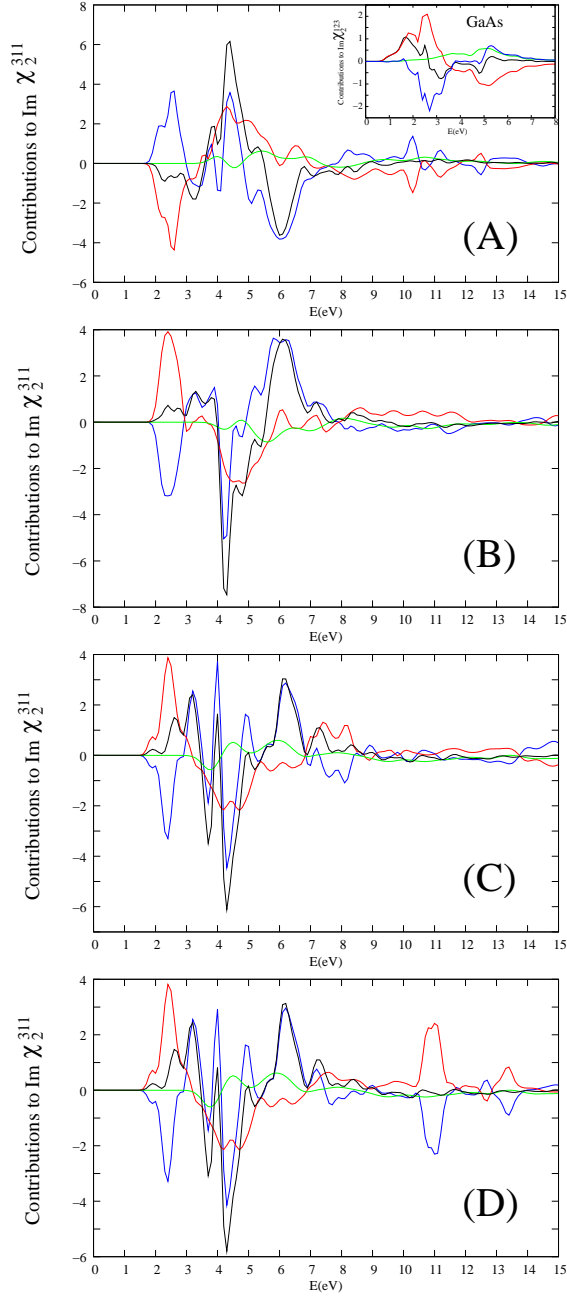}}
\caption{\label{fig:chiim31}(Color online) The different contributions to the imaginary part of the component $\chi^{311}_{2}(-2\omega; \omega,\omega)$. The results (in the unit of $10^{-7}$ esu) of the structure (A), (B), (C), and (D)are exhibited. The results (in the unit of $10^{-6}$ esu) of the $GaAs$ is depicted as the inset of panel (A). The contributions of the inter-band contribution, the intra-band contribution, and the modulation of the inter-band portion by the intra-band transition, are depicted as the blue, the red, and the green lines, respectively; the total is represented by the black line.}
\end{figure}

\begin{figure}[htbp]
\resizebox{0.5\textwidth}{!}{%
  \includegraphics{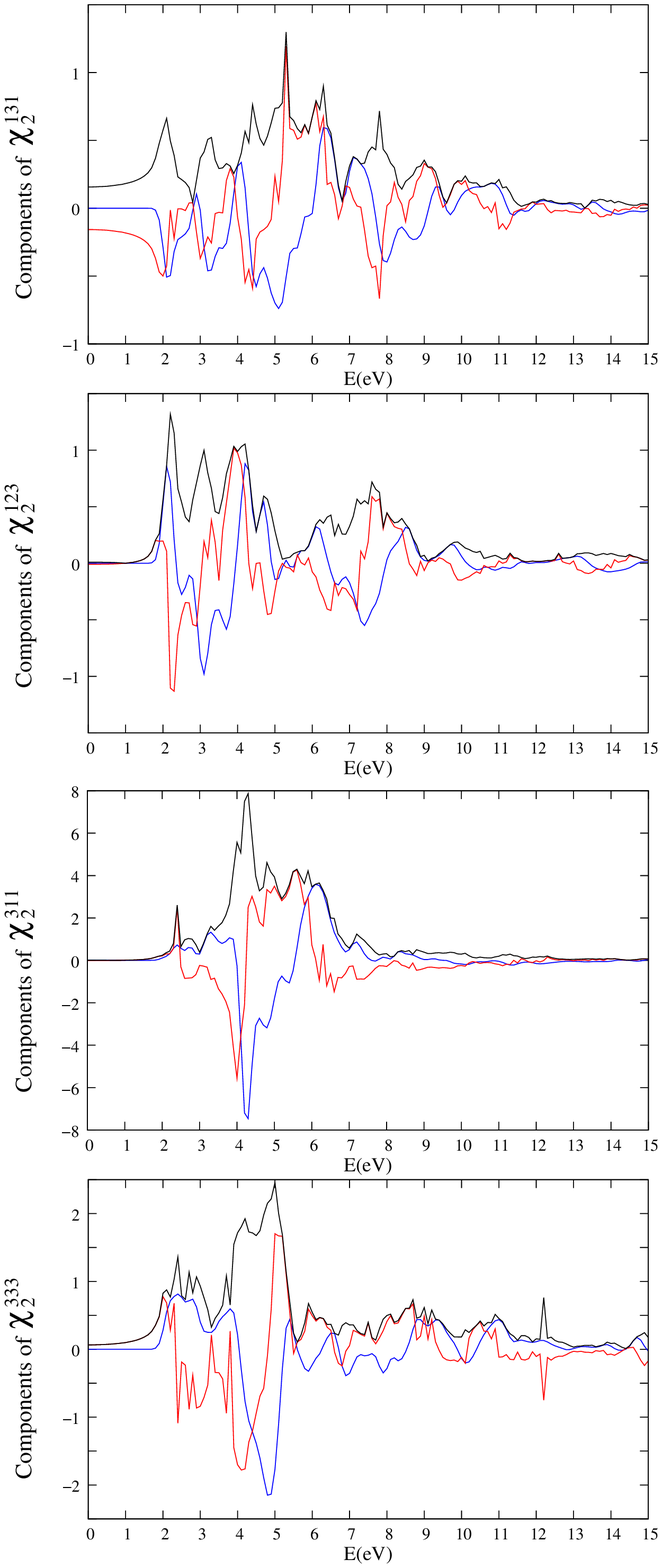}}
\caption{\label{fig:chi}(Color online) The dispersions of all nonzero independent components of the SHG susceptibility tensor (in the unit of $10^{-7}$ esu). The black, the red, and the blue lines depict the absolute, the real, and the imaginary results, respectively.}
\end{figure}

\indent
The results of components $d_{31}$, $d_{33}$, $d_{14}$, and $d_{15}$ at the infinitesimal frequency are tabulated in TABLE~\ref{tab:finited31}, ~\ref{tab:finited33}, ~\ref{tab:finited14}, and ~\ref{tab:finited15}. Basing on the comparison with the experiments to the components $d_{14}$ and $d_{15}$, the resulting magnitude in structure (A) and (B) are comparable to the measured values. Thus, the present study manifests the full bandstructure calculation is valid to yield the results consisting with the experimental data at the infinitesimal frequency for the $\alpha-LiIO_{3}$ crystal. In spite this point infinitively neighboring to the static limit, the contributions associated with the effect of intra-band motion suddenly raise up in contrast to the status at the static limit, reflecting on the differences between the tabulated results of $d_{31}$ in TABLE~\ref{tab:d31}, and ~\ref{tab:finited31}. The cancellation because of the resulting signs of the intra-band and inter-band portions being appositive makes the magnitude of total result at this point much less than that at the static limit. The same trends are also happen to the component $d_{33}$. On the other hand, the aforementioned structural effect is still very remarkable that changes the importance of individual transition mechanism in different structures. To the analogous results in the calculations of $GaAs$, the result $115.8$ pm/V is also near to the corresponding experimental value, $99.8$ pm/V (the value of linear electro-optic coefficient), according to the relationship \cite{Sipe3} $\chi^{xyz}(-\omega;\omega,0)=\chi^{xyz}(-2\omega;\omega,\omega)$ for $\omega$ to be infinitesimal. \\
\indent
The resulting different contributions to the imaginary part of the component $\chi^{311}_{2}(-2\omega; \omega,\omega)$, shown in FIG.~\ref{fig:chiim31}, elucidates the intra-band contribution to be as import as the inter-band one in the dispersion at finite frequencies. The structural effect is still very distinguished according to the resulting dispersions of different structures. In addition, the modification of the pressure effect is appreciable within the low frequency regime but becomes considerable at some high frequency channels to the individual transition mechanism. Besides, the previously predicted double-resonance feature as $\omega$ near to the $E_{g}$ according to the analysis of the bandstructure, \cite{Nie} is reified in the maximal resonant absorption of both inter-band and intra-band transitions exhibited in FIG.~\ref{fig:chiim31}. Especially, this feature very insists regardless of the structural effect. On the other hand, the analogous results in the calculation of $GaAs$, exhibited in FIG.~\ref{fig:chiim31}, very agree with the published results of the references \cite{Sipe3,Rashkeev}.\\
\indent
The dispersions of all nonzero independent components of the SHG susceptibility tensor:
$\chi^{123}_{2}(-2\omega; \omega,\omega)$, $\chi^{131}_{2}(-2\omega;
\omega,\omega)$, $\chi^{311}_{2}(-2\omega; \omega,\omega)$, and
$\chi^{333}_{2}(-2\omega; \omega,\omega)$, are calculated for the
energetically optimal structure, depicted in FIG.~\ref{fig:chi}. The remarkable anisotropic property of the SHG susceptibility of the $\alpha-LiIO_{3}$ crystal reflects on the significant difference among the results of those components.
\section{\label{sec6}SUMMARY}
The full bandstructure calculation is valid to generate correct results of SHG susceptibility to analyze the trends of experimental data at the static limit and at finite frequencies for the $\alpha-LiIO_{3}$ crystal.
According to the agreements with the experiments of the resulting EFG and $E_{g}$, the bandstructure give by the FLAPW calculations is valid, so the non-local defect of DFT-LDA in the semi-conductors can be ignored in the present calculations. The addressed structural effect indeed significantly influences the SHG susceptibility of the $\alpha-LiIO_{3}$ crystal. Thus, it is proposed as one of the major factors causing the anomalous inconsistency of the experimental data collected from published references. On the manipulation of the structural effect by an external pressure, the present simulation manifests this scheme to be effective. The present work resolves the mechanisms of SHG susceptibility of the $\alpha-LiIO_{3}$ crystal, and makes the conclusions: the inter-band transition is determined to dominate the whole SHG susceptibility at the static limit; the effect of the intra-band motion suddenly raise up to be as important as the inter-band transition at finite frequencies. The results on the this respect provide very useful information for the further modification of frequency doubling and associated second order phenomena technologies to the $\alpha-LiIO_{3}$ crystal.
\begin{acknowledgments}
The author thanks Prof. Ding-sheng Wang and his student Dr.
Chun-gang Duan for their assistances in the starting phase of the
present work. This research was financially supported by National
Science Council, R. O. C. (Project No. NSC93-2811-M-001-065).
\end{acknowledgments}

\begin{table*}
\caption{\label{tab:structure} The parameters of the unit cell, the
resulting $V_{zz}$ (in the unit of $10^{21}V/m^{2}$), and the
obtained band gap $E_{g}$ (eV) of the simulated structures. The
coordinates of each atomic species are respectively described as
$Li$-atom$(0,0,z)$, $I$-atom$(1/3,2/3,0)$, and $O$-atom$(x,y,z)$.
The designations of the structures are as followings, (A): the
result of experiment \cite{Rosenzweig}; (B): the result of
experiment \cite{Svensson}; (C): the energetically optimal
structure, and (D): the pertaining distortion to (C).}
\begin{ruledtabular}
\begin{tabular}{cccccccc}
Structure&$a$({\AA})&$c$({\AA})&$Li(z)$&$O(x,y,z)$& $V_{zz}^{Li}$ & $V_{zz}^{I}$ & $E_{g}$\\
\hline
(A)& 5.48150 & 5.170900 & 0.8907 & (0.0936, 0.3440, 0.1698) & 0.036 & -53.244 & 3.735\\
(B)& 5.48169 & 5.172370 & 0.0713 & (0.2468, 0.3419, 0.8377) & 0.057 & -54.016 & 3.934\\
(C)& 5.36816 & 5.019063 & 0.0775 & (0.2655, 0.3387, 0.8327) & 0.044 & -52.514 & 3.297\\
(D)& 5.36816 & 5.019058 & 0.0774 & (0.2656, 0.3391, 0.8328) & 0.044
& -52.584 & 3.304
\end{tabular}
\end{ruledtabular}
\end{table*}
\begin{table*}
\caption{\label{tab:d31} The $d_{31}$ component of SHG
susceptibility tensor (in the unit of pm/V) at the static limit. The
calculated results of the structures (A), (B), (C) and (D) and the
experimental data are tabulated. The 'TOT' labels the result of
equation (\ref{eqn:SHG-st}); the 'Inter' ('Intra') labels the result
of inter-band (intra-band) transition from the equation
(\ref{eqn:SHG-st-inter}) (equation (\ref{eqn:SHG-st-intra})), and
the 'V.H.'('V.E.') labels the result of the visual-hole
(visual-electron) process in the inter-band transition from the
equation (\ref{eqn:SHG-st-inter}). The $\lambda$ labels the operated
wavelength of the cited experimental works.}
\begin{ruledtabular}
\begin{tabular}{cccccccc}
Structure & Intra & Inter & V.H. & V.E. & TOT\\
\hline
(A) & -0.112 & -3.891 & -4.521 & 0.630 & -4.003\\
(B) & 0.032 & 2.671 & 3.330 & -0.659 & 2.703\\
(C) & 0.031 & 4.168 & 3.856 & 0.312 & 4.200\\
(D) & 0.014 & 4.500 & 3.893 & 0.607 & 4.515\\
\hline
Experiment & Result & $\lambda$ (nm)\\
\hline
Exp \cite{Nath} & $-12.2 \pm 1.9$ & 1860\\
Exp \cite{Nash} & $-4.5 \pm 0.6$ & 1856\\
Exp \cite{Campillo} & $-7.54 \pm 1.13$ & -\\
Exp \cite{Jerphagnon} & $-4.96 \pm 0.26$ & 1064.2\\
Exp \cite{Otagurs} & $\pm 10.17 \pm 2.0$ & 514.5\\
Exp \cite{Levine155} & -5.53 & 1064.2\\
Exp \cite{Choy} & -7.215 & -\\
Exp \cite{Choy} & -7.33 & -\\
Exp \cite{Choy} & -6.82 & 1318\\
Exp \cite{Eckardt} & -4.1 & 532\\
Exp \cite{Roberts} & -4.4 & 1064\\
Exp \cite{Borsa} & -4.0 & 488\\
Exp \cite{Alford} & -4.09 & 1064\\
Exp \cite{Alford} & -3.90 & 660\\
Exp \cite{Alford} & -5.23 & 403
\end{tabular}
\end{ruledtabular}
\end{table*}
\begin{table*}
\caption{\label{tab:d33} The $d_{33}$ component of the SHG
susceptibility tensor (in the unit of pm/V) at the static limit.}
\begin{ruledtabular}
\begin{tabular}{cccccccc}
Structure & Intra & Inter & V.H. & V.E. & TOT\\
\hline
(A) & 0.240 & -0.577 & 0.623 & -1.200 & -0.337\\
(B) & -0.171 & 3.444 & 1.724 & 1.720 & 3.272\\
(C) & -0.104 & 1.677 & 0.061 & 1.616 & 1.574\\
(D) & -0.119 & 1.801 & 0.118 & 1.684 & 1.683\\
\hline
Experiment & Result & $\lambda$ (nm)\\
\hline
Exp \cite{Nash} & $-3.6 \pm 1.08$ & 1719\\
Exp \cite{Jerphagnon} & $-5.15 \pm 0.32$ & 1064.2\\
Exp \cite{Choy} & $-6.75 \pm 0.95$ & 1318\\
Exp \cite{Choy} & $-5.54 \pm 0.61$ & 1318\\
Exp \cite{Roberts} & -4.5 & 1064
\end{tabular}
\end{ruledtabular}
\end{table*}
\begin{table*}
\caption{\label{tab:finited31} The $d_{31}$ component of the SHG
susceptibility tensor (in the unit of pm/V) at infinitesimal
frequency. The 'TOT', the 'ter', the 'tra', and the 'mod' label the
contributions of the portion defined respectively from the equations
(\ref{eqn:SHG-FF}), (\ref{eqn:SHG-inter}), (\ref{eqn:SHG-intra}),
and (\ref{eqn:SHG-mod}).}
\begin{ruledtabular}
\begin{tabular}{ccccc}
Structure & ter & tra & mod & TOT\\
\hline
(A) & 4.084 & -7.812 & 3.393 & -0.335\\
(B) & -2.681 & 6.409 & -3.917 & 0.189\\
(C) & -4.105 & 5.152 & -0.168 & 0.879\\
(D) & -5.257 & 6.325 & -0.189 & 0.879
\end{tabular}
\end{ruledtabular}
\end{table*}
\begin{table*}
\caption{\label{tab:finited33} The $d_{33}$ component of the SHG
susceptibility tensor (in the unit of pm/V) at infinitesimal
frequency.}
\begin{ruledtabular}
\begin{tabular}{ccccc}
Structure & ter & tra & mod & TOT\\
\hline
(A) & 0.545 & -1.508 & -0.712 & -1.675\\
(B) & -3.435 & 3.204 & 1.571 & 1.340\\
(C) & -1.655 & 4.800 & 2.367 & 5.512\\
(D) & -1.780 & 4.817 & 2.367 & 5.404
\end{tabular}
\end{ruledtabular}
\end{table*}
\begin{table*}
\caption{\label{tab:finited14} The $d_{14}$ component of the SHG
susceptibility tensor (in the unit of pm/V) at infinitesimal
frequency.}
\begin{ruledtabular}
\begin{tabular}{ccccc}
Structure & ter & tra & mod & TOT\\
\hline
(A) & -0.440 & -1.759 & 1.948 & -0.251\\
(B) & 0.000 & 0.189 & -0.398 & -0.209\\
(C) & -0.042 & 3.540 & 1.990 & 5.488\\
(D) & 0.335 & 3.016 & 2.115 & 5.466\\
\hline Exp. \cite{Singh-up, Okada}& & & & 0.22-0.35 ($\lambda$=
1064.2 nm)
\end{tabular}
\end{ruledtabular}
\end{table*}
\begin{table*}
\caption{\label{tab:finited15} The $d_{15}$ component of the SHG
susceptibility tensor (in the unit of pm/V) at infinitesimal
frequency.}
\begin{ruledtabular}
\begin{tabular}{ccccc}
Structure & ter & tra & mod & TOT\\
\hline
(A) & 3.854 & 1.738 & -2.618 & 2.974\\
(B) & -2.639 & -2.807 & 2.157 & -3.289\\
(C) & -4.147 & 1.299 & 2.115 & -0.733\\
(D) & -4.440 & 1.864 & 1.843 & -0.733\\
\hline Exp. \cite{Pearson} & & & & 5 ($\lambda$= 694.3 nm)
\end{tabular}
\end{ruledtabular}
\end{table*}

\begin{thebibliography}{02}
\bibitem{Kippelen} B. Kippelen, Nature materials {\bf 3}, 841 (2004).
\bibitem{Ferguson} B. Ferguson and X.-C. Zhang, Nature materials {\bf 1}, 26 (2002).
\bibitem{Buesener} H. Buesener, A. Renn, M. Brieger, F. von Moers, and A. Hese, Appl. Phys. B {\bf 39}, 77 (1986).
\bibitem{Nath} G. Nath and S. Haussuhl, Appl. Phys. Lett. {\bf 14}, 154 (1969).
\bibitem{Nash} F. R. Nash, J. G. Bergman, G. D. Boyd, and E. H. Turner, J. of Appl. Phys. {\bf 40}, 5201 (1969).
\bibitem{Campillo} F. R. Campillo and C. L. Tang, Appl. Phys. Lett. {\bf 16}, 424 (1970).
\bibitem{Jerphagnon} J. Jerphagnon, Appl. Phys. Lett. {\bf 16}, 298 (1970).
\bibitem{Otagurs} W. S. Otagurs, E. Weimer-Avnear, and S. P. S. Porto, Appl. Phys. Lett. {\bf 18}, 499 (1971).
\bibitem{Levine155} B.F. Levine and C. G. Bethea, Appl. Phys. Lett. {\bf 20}, 272 (1972).
\bibitem{Choy} Michael M. Choy and Robert L. Byer, Phys. Rev. B {\bf 14}, 1693 (1976).
\bibitem{Eckardt} Robert C. Eckardt, Hisashi Masuda, Yuan Xuan Fan, and Robert L. Byer, IEEE J. Quantum Electron. {\bf 26}, 922 (1990).
\bibitem{Roberts} David A. Roberts, IEEE J. Quantum Electron. {\bf 28}, 2057 (1992).
\bibitem{Borsa} Giuliana Borsa, Stefania Castelletto, Aldo Godone, Carlo Novero, and Luisa Rastello, Opt. Rev. {\bf 4}, 484 (1997).
\bibitem{Alford} William J. Alford and Arlee V. Smith, J. Opt. Soc. Am. B, {\bf 18}, 524 (2001)
\bibitem{Singh-up} S. Singh, J. R. Potopowicz, W. A. Bonner and L. G. Van Uitert,
unpublished work was quoted in the handbook \cite{Singh}.
\bibitem{Okada} M. Okada and S. Ieiri, Phys. Lett., {\bf 34A}, 63 (1971).
\bibitem{Pearson} J. E. Pearson, G. A. Evans, and A. Yariv, Opt. Commun. {\bf 4}, 366 (1972).
\bibitem{Liu} Yung S. Liu, Appl. Phys. Lett. {\bf 31}, 187 (1977).
\bibitem{Kato} K. Kato, IEEE J. Quantum Electron. {\bf 21}, 119 (1985).
\bibitem{Teyssier1} J. Teyssier, R. Le Dantec, C. Galez, Y. Mugnier, A. Vrain, J. Bouillot, and J.-C. Plenet, Proc. SPIE {\bf 5946}, 59460J (2005).
\bibitem{Teyssier2} J. Teyssier, R. Le Dantec, C. Galez, Y. Mugnier, J. Bouillot, and J.-C. Plenet, Applied Phys. Lett. {\bf 85}, 710 (2004).
\bibitem{Singh}S. Singh, {\it Handbook of Laser Science and Technology}, vol. {\bf III}, edited by M. J. Weber (CRC Press, Boca Raton, FL, 1986); references therein.
\bibitem{Svensson} C. Svensson, J. Albertsson, R. Liminga, \AA. Kvick, and S. C. Abrahams, J. Chem. Phys. {\bf 78}, 7343 (1983).
\bibitem{Rosenzweig} Abraham Rosenzweig and Bruno Morosin, Acta Cryst. {\bf 20}, 758 (1966).
\bibitem{Emiralieb} A. Emiralieb et al, Kristallografiya {\bf 18}, 1177 (1973).
\bibitem{Gubler} U. Gubler and C. Bosshard, Nature materials {\bf 1}, 209 (2002).
\bibitem{Nie} Yung-mau Nie, the article submitted to the Appl. Phys. A.
\bibitem{Sipe1} J. E. Sipe and Ed Ghahramani, Phys. Rev. B {\bf 48}, 11705 (1993).
\bibitem{Sipe2} Claudio Aversa and J. E. Sipe, Phys. Rev. B {\bf 52}, 14636 (1995).
\bibitem{Sipe3} James L. P. Hughes and J. E. Sipe, Phys. Rev. B {\bf 53}, 10751 (1996).
\bibitem{Sipe4} James L. P. Hughes, Y. Wang and J. E. Sipe, Phys. Rev. B {\bf 55}, 13630 (1997).
\bibitem{Sipe5} J. E. Sipe and A. I. Shkrebtii, Phys. Rev. B {\bf 61}, 5337 (2000).
\bibitem{Aspnes} D. E. Aspnes, Phys. Rev. B {\bf 6}, 4648 (1972).
\bibitem{Levine} Zachary H. Levine and Douglas C. Allan, Phys. Rev. Lett. {\bf 66}, 41 (1991); Phys. Rev. B {\bf 44}, 12781 (1991); Zachary H. Levine, Phys. Rev. B {\bf 49}, 4532 (1994).
\bibitem{Lax} Melvin Lax, {\it SYMMETRY PRINCIPLES IN SOLID STATE AND MOLECULAR PHYSICS} (Dover, New York, 2001).
\bibitem{Corso} A. D. Corso, F. Mauri, and A. Rubio, Phys. Rev. B {\bf 53}, 15638 (1996), and references therein.
\bibitem{Levine1970} B. F. Levine, Phys. Rev. Lett. {\bf 25}, 440 (1970).
\bibitem{Vanderbilt} D. Vanderbilt and R. Resta, "Quantum electrostatics of insulators: Polarization, Wannier functions, and electric fields," in \textit{Conceptual foundations of materials properties: A standard model for calculation of ground- and excited-state properties}, S.G. Louie and M.L. Cohen, eds. (Elsevier, The Netherlands, 2006), pp. 139-163.
\bibitem{Rashkeev} Sergey N. Rashkeev, Walter R. L. Lambrecht, and Benjamin Segall, Phys. Rev. B {\bf 57}, 3905 (1998).
\bibitem{Kleinman}D. A. Kleinman, Phys. Rev. {\bf 126}, 1977 (1962).
\bibitem{Franken} P. A. Franken and J. F. Ward, Rev. of mod. Phys. {\bf 35}, 23 (1963).
\bibitem{Marques} M. A. L. Marques and E. K. U. Gross, Annual Rev. of Phys. Chem. {\bf 55}, 427 (2004), and references therin.
\bibitem{Aulbur} W. G. Aulbur, L. J\"{o}nsson, and J. W. Wilkins, Phys. Rev. B {\bf 54}, 8540 (1996).
\bibitem{Lambrecht} W. R. L. Lambrecht and B. Segall, Phys. Rev. B {\bf 40}, 7793 (1989).
\bibitem{Levin43} Z. H. Levine and D. C. Allan, Phys. Rev. B {\bf 43}, 4187 (1991).
\bibitem{Nastos} F. Nastos, B. Olejnik, K. Schwarz, and J. E. Sipe, Phys. Rev. B {\bf 72}, 45223 (2005).
\bibitem{PBE} J. P. Perdew, S. Burke, and M. Ernzerhof, Phys. Rev. Lett. {\bf 77}, 3865 (1996).
\bibitem{Blochl1} P. E. Bl\"{o}chl, Phys. Rev. B {\bf 50}, 17953 (1994).
\bibitem{Kresse1} G. Kresse and D. Joubert, Phys. Rev. B {\bf 59}, 1758 (1999).
\bibitem{Blochl} P. E. Bl\"{o}chl, O. Jepsen and O. K. Andersen, Phys. Rev. B {\bf 49}, 16223 (1994).
\bibitem{Kresse2} G. Kresse and J. Furthmuller, J. Comput. Mater. Sci. {\bf 6}, 15 (1996).
\bibitem{Kresse3} G. Kresse and J. Furthmüller, Phys. Rev. B {\bf 54}, 11169 (1996).
\bibitem{Blaha1} P. Blaha, K. Schwarz, and P. Herzig, Phys. Rev. Lett. {\bf 54}, 1192 (1985).
\bibitem{Blaha2} P. Blaha, K. Schwarz, and P. H. Dederichs, Phys. Rev. B {\bf 37}, 2792 (1988).
\bibitem{Blaha3} Helena M. Petrilli, Peter E. Bl\"{o}chl, P. Blaha, and K. Schwarz, Phys. Rev. B {\bf 57}, 14690 (1998).
\bibitem{Madsen} Georg K. H. Madsen, Peter Blaha, Karlheinz Schwarz, Elisabeth Sjostedt, and Lars Nordstrom, Phys. Rev. B {\bf 64}, 195134 (2001).
\bibitem{Monkhorst} H. J. Monkhorst and J. D. Pack, Phys. Rev. B {\bf 13}, 5188 (1976).
\bibitem{Regel} L. L. Regel', Z. B. Perekalina, A. I. Baranov, B. V. Shchepetil'nikov, and N. A. Baturin, Sov. Phys. Crystallogr. {\bf 32}, 862 (1988).
\bibitem{Xu} Jingjun Xu, Xuefeng Yue, and Romano A. Rupp, Phys. Rev. B {\bf 54}, 16618 (1996).
\bibitem{Galez} C. Galez, Y. Mugnier, J. Bouillot and C. Rosso, Optical Materials {\bf 19}, 33 (2002).
\bibitem{Gaffar} M. A. Gaffar, A. Abu El-Fadl, J. of Phys. and Chem. of Solids {\bf 60}, 1633 (1999).
\bibitem{Christiansen} {\it Hyperfine Interaction of Radioactive Nuclei}, edited by J. Christiansen, Topics in current Physics (Springer, New York, 1983), Vol. 31, p. 133.
\bibitem{Baisa} D. F. Ba\u{i}sa and S. V. Mal'tsev, Sov. Phys. Solid State {\bf 29}, 516 (1987).
\bibitem{Pyykko} P. Pyykk\u{o}, Z. Naturforsch. {\bf 47a}, 189 (1992).
\bibitem{Minamisono} T. Minamisono, T. Ohtsubo, Y. Nakayama, S. Fukuda, T. Izumikawa, M. Tanigaki, M. Matsui, S. Takeda, n. Nakamura, M. Fukuda, K. Matsuta and Y. Nojiri, Hyperfine Interactions {\bf 78}, 159 (1993).
\bibitem{Voelk} H.-G. Voelk and D. Fick, Nuclear Phys., {\bf A530}, 475 (1991).
\end{thebibliography}
\end{document}